\documentclass[aps,preprint,floatfix]{revtex4-1}
\usepackage{bm}
\usepackage{amsmath}
\usepackage{epsfig}
\usepackage{rotating}
\usepackage{graphicx}
\def\lazz{\mathrel{\mathchoice {\vcenter{\offinterlineskip\halign{\hfil
$\displaystyle##$\hfil\cr<\cr\sim\cr}}}
{\vcenter{\offinterlineskip\halign{\hfil$\textstyle##$\hfil\cr<\cr\sim\cr}}}
{\vcenter{\offinterlineskip\halign{
\hfil$\scriptstyle##$\hfil\cr<\cr\sim\cr}}}
{\vcenter{\offinterlineskip\halign{\hfil$\scriptscriptstyle##
$\hfil\cr<\cr\sim\cr}}}}}

\def\be{\begin{equation}}
\def\lan{\left\langle}
\def\ran{\right\rangle}
\def\ee{\end{equation}}
\def\barr{\begin{array}}
\def\earr{\end{array}}

\def\l{\left}
\def\r{\right}
\def\dis{\displaystyle}
\def\ed{\end{document}}

\def\co{{\cal O}}
\def\ch{{\cal H}}
\def\can{{\cal N}}

\def\cg{{\cal G}}
\def\ce{{\cal E}}

\def\cn{{\bf n}}

\def\tmp{\widetilde{m_p}}
\def\tmn{\widetilde{m_n}}

\def\wm{{\widetilde {m}}}

\begin{document}

\title{Statistical shell model for neutrinoless double
$\beta$-decay nuclear transition matrix elements: Results for $^{76}$Ge, $^{82}$Se, $^{100}$Mo, $^{124}$Sn,
$^{130}$Te and $^{136}$Xe}

\author{V.K.B. Kota$^{1}$\footnote{Corresponding author,
{\it E-mail address:}
vkbkota@prl.res.in}, R. Sahu$^{2}$}

\affiliation{$^1$Physical Research  Laboratory, Ahmedabad 380 009, India \\ $^2$NIST University, Institute Park, Berhampur 761008, India}

\begin{abstract}

Statistical shell model (also called spectral distribution method or statistical spectroscopy method) based on random matrix theory and spherical shell model gives a theory for calculating neutrinoless double beta decay nuclear transition matrix elements (NDBD-NTME). This theory is briefly described and then applied to $^{76}$Ge, $^{82}$Se, $^{100}$Mo, $^{124}$Sn,$^{130}$Te and $^{136}$Xe NDBD-NTME. In these calculations, the Bethe's spin-cutoff factor and a bivariate correlation coefficient are varied in a range dictated by random matrix theory and trace propagation. The calculated NDBD-NTME are compared with the results from several other models as available in literature. The statistical shell model results are in general a factor 2 smaller compared to those from the spherical shell model. 

\end{abstract}

\maketitle

\section{Introduction}

Neutrinoless double beta decay ($0\nu \beta \beta$ or NDBD) which involves emission of two electrons without the accompanying neutrinos and which violates 
lepton number conservation has been an important and challenging problem both for the experimentalists and theoreticians.  Recent neutrino oscillation 
experiments have demonstrated that neutrinos have mass
\cite{Kamok,SNO-1,SNO-2}. The observation of $0\nu \beta \beta$ decay is expected to provide information regarding  the absolute  neutrino mass which is, as yet, not known. As a result, experimental programs to observe this decay
have  been initiated at different laboratories across the globe and the latest results, upto year 2023, giving lower bound on NDBD half-lives, from various experiments are reviewed in \cite{Barb-23}. For example, the most recent results for $0\nu \beta \beta$ decay of $^{136}$Xe have been reported by KamLand-Zen  collaboration \cite{Kamland}. They give a lower limit of $2.3\times 10^{26}$ yr for the half-life (other experiment for
$^{136}$Xe is EXO-200 \cite{Exo-200}). Further, GERDA experiment \cite{Gerda} for $^{76}$Ge give a lower limit of $1.8\times 10^{26}$ yr for the half-life (other experiment for $^{76}$Ge is Majorana \cite{Major}). Other nuclei of considerable interest are $^{48}$Ca \cite{Candles}, $^{82}$Se \cite{Cupid0,Nemo3}, $^{100}$Mo \cite{CUPID-Mo,Amore}, $^{116}$Cd \cite{AURORA} and $^{128,130}$Te \cite{Curoe}. In the present paper we consider $^{76}$Ge, $^{82}$Se. $^{100}$Mo, $^{130}$Te and $^{136}$Xe. Added also is $^{124}$Sn as this is of interest for India based Neutrino Observatory (INO) \cite{Sn124-exp}. It is important that the experimental sensitivity of $0\nu \beta \beta$ decay half life
(in Yr) has reached $\sim 10^{26}$ (90\% C.L.) and in the future experiments (LEGEND for $^{76}$Ge, CUPID and AMoRE for $^{100}$Mo, SNO+ for $^{130}$Te and nEXO and KamLand2-Zen for $^{136}$Xe) the sensitivity
will be increased to $10^{27} - 10^{28}$ so that $\lan m_\nu\ran \lazz 10-20$ meV; for more details on future experiments see \cite{Barb-23,Amore} and references therein.
  
Turning to theory, nuclear transition matrix elements (NTME) are the essential ingredient for extracting the neutrino mass from the measured half lives \cite{Vogel}. There has been
considerable effort to obtain NTME for various candidate nuclei and they are calculated using a variety of nuclear models and also each of them by many different groups. Commonly used are one of the following four models: (i) nuclear shell model (SM) by Horoi, Poves, Coraggio and others \cite{th-sm1,th-sm2,th-sm3,Sn124,Poves,th-sm4}; (ii) quasi-particle random phase approximation (QRPA) and its variants by Faessler, Simkovic, Suhonen and others \cite{th-qrpa1,th-qrpa2,th-qrpa3};  (iii) proton-neutron interacting boson model (pnIBM or IBM-2) by Iachello and collaborators \cite{th-iac1,th-iac2}; (iv) particle number and angular momentum projection including
configuration mixing within the generating coordinate method framework, i.e energy density functional method (EDF) by Rodriguez, Song and others \cite{th-edf1,th-edf2}. In addition, there is a beginning in employing ab-initio methods such as no-core shell model and quantum Monte-Carlo method (at present only for light nuclei ab-initio methods are applied) \cite{th-ab1,th-ab2}. A detailed comparative study of  the results from these models is given in a recent review \cite{Rmp-2023}; see also \cite{Amore}. For further details see Section \ref{sec4}. In addition to (i)-(v), more recently the  so called deformed shell model (DSM) based on Hartree-Fock single particle states has been used for some of the
candidate nuclei in the A=60-90 region \cite{KS} and similarly, projected Hartree-Fock-Bogoliubov method (PHFB) with pairing plus quadrupole-quadrupole interaction is also used \cite{RR}. The NTME obtained from all these theoretical calculations are found to be typically in the range of 2 to 6 \cite{Rmp-2023}. Yet another approach to obtain NTME is to
employ the statistical shell model method. 

Statistical shell model (SSM) is often called spectral distribution method (SDM) or statistical spectroscopy (SS) method \cite{CFT,DFW-1,DFW-2,Wong,FKPT,Gomez,KH}. The word SSM was used first by Zelevinsky \cite{Zel-1,Zel-2}. It is important to mention that SSM is used successfully for calculating nuclear level densities in $2s1d$ and $2p1f$ shell nuclei \cite{Zel-1,Zel-2,KM-96,Zel-3,lvlden}. Also, in the past SSM/SDM was used for calculating transition matrix elements needed for example for: (i) $\beta$-decay rates of 
$fp$-shell nuclei for presupernova stars \cite{stren1,stren2}; (ii) electromagnetic transition strength sums \cite{stren3}; (iii) single nucleon transfer strengths \cite{stren4}; (iv) deriving bounds on parity breaking term in nucleon-nucleon interaction \cite{stren5} and so on. It is important to note that SSM/SDM has its basis in random matrix theory and in particular on the operation of embedded Gaussian orthogonal ensemble of random matrices (EGOE) in nuclear shell model spaces. See \cite{MF,FKPT,Br-81,BW-1,BW-2,Ko-book,KC-2018} for details of EGOE. Following all the developments in SSM/SDM, We are prompted to apply this statistical theory for the calculation of NDBD-NTME. First set of calculations are carried out for NDBD nuclei $^{76}$Ge, $^{82}$Se, $^{100}$Mo, $^{124}$Sn, $^{130}$Te and $^{136}$Xe. The purpose of this article is to describe these results obtained from SSM. Let us mention that the results for $^{76}$Ge, $^{82}$Se, $^{130}$Te and $^{136}$Xe were reported in two conference proceedings \cite{KH-2016,Ko-2017} and briefly also in a review on EGOE applications in nuclei \cite{KC-2018}. We will now give a preview.

Section \ref{sec2} gives a brief discussion of the relation
between neutrino mass and NTME and then describe briefly the structure of the NDBD transition operator. Section \ref{sec3} gives the SSM formalism, with all essential formulas, for calculating NDBD-NTME. In Section \ref{sec4}, we will present the SSM results for $^{76}$Ge, $^{82}$Se, $^{124}$Sn, $^{100}$Mo, $^{130}$Te and $^{136}$Xe. We will compare all the results with those from other nuclear models. Finally, Section \ref{sec5} gives conclusions and future outlook. 

\section{Neutrinoless double beta decay nuclear transition matrix elements}
\label{sec2}

In $0\nu \beta \beta$, the half-life for the 0$^+_i$ ground state (gs) of a
initial even-even nucleus decaying to the 0$^+_f$ gs of the final even-even nucleus
is  given by \cite{Vogel}
\be
\l[ T_{1/2}^{0\nu}(0^+_i \to 0^+_f) \r]^{-1} =  G^{0\nu}
\l| M^{0\nu} (0^+)\r|^2 \l(\dis\frac{\lan m_\nu \ran}{m_e}\r)^2 \;,
\label{eq.dbd1}
\ee
where $\lan m_\nu \ran$ is the effective neutrino mass (a combination of
neutrino mass eigenvalues and also involving the neutrino mixing matrix).  The
$G^{0\nu}$ is a phase space integral  (kinematical factor); tabulations for 
$G^{0\nu}$ are available in literature. The $M^{0\nu}$ represents NTME of the NDBD  transition
operator and it is a sum of a Gamow-Teller like ($M_{GT}$), Fermi like ($M_F$)
and tensor ($M_T$) two-body operators. Since it is well known that the  tensor
part contributes only up to 10\% of the matrix elements,  we will neglect the
tensor part. Then, from the closure approximation which is well justified for
NDBD, we have
\be
\barr{rcl}
M^{0\nu} (0^+) & = & M^{0\nu}_{GT} (0^+) - \dis\frac{g_V^2}{g_A^2} 
M^{0\nu}_{F} (0^+) = \lan 0^+_f \mid\mid \co(2:0\nu) \mid\mid 0^+_i \ran \;,
\\
\co(2:0\nu) & = & \dis\sum_{a,b} \ch(r_{ab}, \overline{E}) 
\tau_a^+ \tau_b^+ \l( \sigma_a \cdot \sigma_b- \dis\frac{g_V^2}{g_A^2} \r)
\;.
\earr \label{eq.dbd2}
\ee
As seen from Eq. (\ref{eq.dbd2}),  NDBD half-lives are generated by the two-body
transition operator $\co(2:0\nu)$; note that $a,b$ label  nucleons. The $g_A$
and $g_V$ are the weak axial-vector and vector coupling constants.  The
$\ch(r_{ab}, \overline{E})$ in Eq. (\ref{eq.dbd2}) is called the `neutrino
potential'. Here, $\overline{E}$ is the average energy of the virtual
intermediate states used in the closure approximation. The form given by Eq.
(\ref{eq.dbd2}) is justified {\it only if the exchange of the light Majorana
neutrino is indeed the mechanism responsible for the NDBD}. With the phase space
factors fairly well known, all one needs are NTME $|M^{0\nu} (0^+)|=\l|\lan
0^+_f \mid\mid \co(2:0\nu) \mid\mid 0^+_i \ran\r|$. Then, measuring the 
half-lives makes  it possible to deduce neutrino mass using Eq. (\ref{eq.dbd1}).

The neutrino potential is of the form $\ch(r_{ab}, \overline{E}) =
[R/r_{ab}]\,\Phi(r_{ab},\overline{E})$ where $R$ in fm units is the nuclear
radius and similarly  $r_{ab}$ is in fm units. A simpler form for the function
$\Phi$, involving sine and cosine integrals, as given in \cite{Vogel} and
employed in \cite{KS}, is used in the present work. It is useful to note that
$\Phi(r_{ab},\overline{E}) \sim  \exp ({-\frac{3}{2} \frac{\overline{E}}{\hbar
c} r_{ab}})$.   The effects of short-range correlations in the wavefunctions are
usually taken into  account by multiplying the wavefunction by the Jastrow
function $[1 - \gamma_3e^{-\gamma_1 r_{ab}^2}  ( 1 - \gamma_2 r_{ab}^2 )]$; $(\gamma_1,\gamma_2,\gamma_3)$ are free parameters.
There are other approaches for taking into account the short range
correlations  but they are not considered here. Now, keeping the wavefunctions
unaltered, the Jastrow  function can be incorporated into $\ch(r_{ab},
\overline{E})$ giving an effective $\ch_{eff}(r_{ab}, \overline{E})$,
\be
\ch(r_{ab}, \overline{E}) \to 
\ch_{eff}(r_{ab}, \overline{E}) = \ch(r_{ab}, \overline{E}) 
[ 1 - \gamma_3 \;e^{-\gamma_1 \; r_{ab}^2} ( 1 - \gamma_2 \; r_{ab}^2 ) 
]^2 \;.
\label{eq.dbd3b}
\ee 
In all the calculations presented in this article, the various parameters in Eq. (3) are chosen  to be (i) $R=1.2A^{1/3}$ fm; (ii) $b=1.003A^{1/6}$ fm; (iii) $\overline{E}=1.12A^{1/2}$ MeV; (iv) $g_A/g_V=1$ (quenched); (v) $\gamma_1=1.1$ fm$^{-2}$, $\gamma_2=0.68$ fm$^{-2}$ and $\gamma_3=1$. These $(\gamma_1,\gamma_2,\gamma_3)$ values are Miller-Spencer Jastrow correlation parameters and these are employed in the present work. Other choices used in
literature are CD-Bonn with $(\gamma_1,\gamma_2,\gamma_3)$=(1.52 fm$^{-2}$ ,1.88 fm$^{-2}$ ,0.46) and
AV18 with $(\gamma_1,\gamma_2,\gamma_3)$=(1.59fm$^{-2}$  ,1.45 fm$^{-2}$ ,0.92).

Let us say that for the nuclei under consideration, protons are in the single
particle (sp) orbits $j^p$ and similarly neutrons in $j^n$. Using the usual
assumption that the radial part of the sp states are those of the harmonic
oscillator, the proton sp states are completely specified by
($\cn^p,\ell^p,j^p$) with $\cn^p$ denoting oscillator radial quantum number so
that for a oscillator shell $\can^p$, $2\cn^p+\ell^p=\can^p$. Similarly, the
neutron sp states are ($\cn^n,\ell^n,j^n$). In terms of the creation
($a^\dagger$) and annihilation ($a$) operators, normalized two-particle
(antisymmetrized) creation operator $A^J_\mu(j_1j_2) = 
(1+\delta_{j_1j_2})^{-1/2}  (a^\dagger_{j_1}a^\dagger_{j_2})^J_\mu$ and then
$A^J_\mu \l|0\ran = \l|(j_1 j_2)J \mu\ran$ represents a normalized  
two-particle state. At this stage, it is important to emphasize that we are
considering only $0^+$ to $0^+$ transitions in $0 \nu \beta \beta$ and therefore
only the $J$ scalar part of $\co(2:0\nu)$ will contribute to $M^{0\nu}$. With
this, the NDBD  transition operator  can be written as,
\be
\co(2:0\nu) = \dis\sum_{j_1^p \geq j_2^p;j_3^n \geq j_4^n;J} \co_{j_1^p \, 
j_2^p; j_3^n \, j_4^n}^J (0\nu) \dis\sum_\mu A^J_\mu(j_1^pj_2^p) \l\{
A^J_\mu(j_3^nj_4^n) \r\}^\dagger \;.
\label{eq.dbd4}
\ee  
Here, $\co_{j_1^p \, j_2^p; j_3^n \, j_4^n}^J (0\nu) = \lan (j_1^p \,  j_2^p) J
M \mid \co(2:0\nu)  \mid (j_3^n \, j_4^n) J M \ran_a$ are two-body matrix
elements (TBME) and \lq{$a$}\rq in the suffix denotes antisymmetrized two-particle
wavefunctions; $J$ is even  for $j_1=j_2$  or $j_3=j_4$. The TBME in this work are obtained by using the standard approach based on Brody-Moshinisky  brackets and Talmi integrals \cite{Br-Ta}. An alternative approach is based on Horie method \cite{Horie} and this is used for example by Iachello et al. \cite{th-iac1}.  

\section{Statistical shell model method for nuclear transition matrix elements}
\label{sec3}

Let us consider shell model sp orbits $j^p_1,  j^p_2, \ldots, j^p_r$ with $m_p$
protons distributed in them. Similarly, $m_n$ neutrons are distributed in
$j^n_1, j^n_2, \ldots, j^n_s$ orbits. Note that the number of proton sp states for a $j_i^p$ orbit is $N_i^p=(2j^p_i+1)$ and then the total number of proton sp states is given by $N_p=\sum_i N^p_i$. Similarly,
$N_i^n=(2j^n_i+1)$ and $N_n=\sum_i N^n_i$. Going further, each distribution of the $m_p$ number of protons in the proton orbits will give a proton configuration
$\tmp=[m_p^1, m_p^2, \ldots, m_p^r]$ where $m_p^i$ is number of protons in the
orbit $j_i^p$ with $\sum_{i=1}^r\,m_p^i=m_p$. Similarly, the neutron
configurations are $\tmn=[m_n^1, m_n^2, \ldots, m_n^s]$ where $m_n^i$ is number
of neutrons in the orbit $j_i^n$ with $\sum_{i=1}^s\,m_n^i=m_n$. With these,
$(\tmp, \tmn)$'s denote proton-neutron configurations.  The nuclear effective
Hamiltonian is one plus two-body, $H=h(1)+V(2)$ with the one-body part $h(1)$ defined by the single particle energies (spe) and the two-body part $V(2)$ by a set of TBME. The total state density $I^{(m_p,m_n)}(E)$, giving number of states (this will count $(2J+1)$ factor) per unit energy interval at a given
energy $E$, is a sum of the partial densities (or local density of states) defined over $(\tmp, \tmn)$ and then \cite{CFT,Wong,KH} 
\be
I^{(m_p, m_n)}(E) = \dis\sum_{(\tmp,\tmn)}\,I_{\cg}^{(\tmp, \tmn)}(E) = \dis\sum_{(\tmp,\tmn)}\,d(\tmp,\tmn)\,\rho_{\cg}^{(\tmp, \tmn)}(E) \;.
\label{eq.ndbd1}
\ee
Here, $d(\tmp,\tmn) = \prod_i \binom{N_i^p}{m_p^i}\,\prod_j \binom{N_j^n}{m_n^j}$  is the dimension of the configuration $(\tmp, \tmn)$, $\cg$ denotes Gaussian and the partial density $\rho_{\cg}^{(\tmp, \tmn)}(E)$ is normalized to unity. Note that the $I^{(m_p, m_n)}(E)$ is normalized to the total dimension $d(m_p,m_n) = \binom{N_p}{m_p} \binom{N_n}{m_n}$.
For strong enough two-body
interactions (this is valid for nuclear interactions \cite{KH}), the operation 
of embedded GOE of one plus two-body interactions [EGOE(1+2)] will lead to 
Gaussian form for the partial densities $\rho^{(\tmp, \tmn)}(E)$ \cite{Ko-01,Ko-book} and then we have the Gaussian  partial densities as shown in Eq. (\ref{eq.ndbd1}). The Gaussian partial densities
are defined by the energy centroids $E_c(\tmp, \tmn)=\lan H \ran^{(\tmp, \tmn)}$ and
variances $\sigma^2(\tmp, \tmn) = \lan H^2  \ran^{(\tmp, \tmn)} - [E_c(\tmp,\tmn)]^2$,
\be
\rho_{\cg}^{(\tmp, \tmn)}(E) = \dis\frac{1}{\dis\sqrt{2\pi}\;\sigma(\tmp, \tmn)}\;\exp-\l\{\l[E-E_c(\tmp, \tmn)\r]^2 /2\,\sigma^2(\tmp, \tmn)\r\}\;.
\label{eq.ndbd2}
\ee 
Exact formulas for $E_c(\tmp, \tmn)$ and $\sigma^2(\tmp, \tmn)$ in terms of spe and TBME, follow easily from trace propagation
methods \cite{CFT,KH}. In practical applications to nuclei, we need level densities and therefore Eq. (\ref{eq.ndbd1})
has to be applied in fixed-$J$ spaces \cite{Zel-1,Zel-2,Zel-3} or an approximate $J$ projection of $\rho_{\cg}^{(\tmp, \tmn)}(E)$   has to be carried out \cite{KM-96,Fr-06}. We will return to this later.

Our interest is in calculating NTME
$$
\l|\lan 0^+_f \mid\mid \co(2:0\nu) \mid\mid 0^+_i\ran\r|^2\;.
$$
Given a transition operator $\co$, operation of EGOE(1+2) in shell model spaces gives a statistical theory for squares of matrix elements of $\co$ in $H$ eigenstates \cite{FKPT,KH}. 
Let us say that $E_i$ are energies of the initial nucleus and $E_f$ are energies of the final nucleus involved in the neutrinoless double beta decay. Also, denote the proton-neutron configuration in the shell model space of the initial nucleus by $\wm_i$ and similarly for the final nucleus we have
$\wm_f$. Then \cite{KH-2016,KC-2018},
\be
\barr{l}
\l|\lan E_f \mid \co \mid E_i \ran\r|^2 = \dis\sum_{\wm_i,\wm_f} 
\dis\frac{
I^{\wm_i}_{\cg}(E_i) I^{\wm_f}_{\cg}(E_f)}{I^{m_i}(E_i) I^{m_f}(E_f)}
\l|\lan \wm_f \mid \co \mid \wm_i\ran\r|^2 \\ 
\times \dis\frac{\rho^H_{\co : biv-\cg}(E_i , E_f , 
\ce_{\co,H}(\wm_i), \ce_{\co,H}(\wm_f),
\sigma_{\co,H}(\wm_i), \sigma_{\co,H}(\wm_f), \zeta_{\co,H}(\wm_i , \wm_f)}{\rho_{\cg}^{\wm_i}(E_i)
\rho_{\cg}^{\wm_f}(E_f)}\;; \\ 
\l|\lan \wm_f \mid \co \mid \wm_i\ran\r|^2 = \l[d(\wm_i) d(\wm_f) \r]^{-1}\;
\dis\sum_{\alpha , \beta} 
\l|\lan \wm_f, \alpha \mid \co \mid \wm_i, \beta \ran\r|^2\;.
\earr \label{eq.ndbd3}
\ee
Eqs. (\ref{eq.ndbd1}) - (\ref{eq.ndbd3}) will allow one to use SSM for the calculation of
NTME $M^{0\nu}$. In Eq. (\ref{eq.ndbd3}), $\rho^H_{\co : biv-\cg}$ is a bivariate Gaussian and this needs five parameters: the marginal centroids
$\ce_{\co,H}(\wm_i)$ and $\ce_{\co,H}(\wm_f)$, marginal variances $\sigma^2_{\co,H}(\wm_i)$ and $\sigma^2_{\co,H}(\wm_f)$ and the correlation coefficient $\zeta_{\co,H}(\wm_i,\wm_f)$. The marginal centroids and variances in Eq. (\ref{eq.ndbd3}) are approximated, following random matrix theory \cite{Ko-15}, to the  corresponding
state density centroids and variances giving
$\ce_{\co,H}(\wm_r) \approx E_c((\tmp, \tmn)_r) =  \lan H \ran^{(\tmp,\tmn)_r}$ and $\sigma_{\co,H}^2(\wm_r) \approx  \sigma^2((\tmp,
\tmn)_r) = \lan H^2 \ran^{(\tmp, \tmn)_r} - \l[E_c((\tmp, \tmn)_r) \r]^2$; $r=i,f$.  For the correlation coefficient $\zeta$, there is not yet any valid form involving configurations. A plausible way forward, as used in \cite{KH-2016,KC-2018}, is to calculate $\zeta$ as a
function of $(m_p,m_n)$ using random matrix theory given in \cite{Ko-15}. For nuclei of interest, using the numerical results in Table 2 of \cite{Ko-15}, it is seen that $\zeta \sim 0.6-0.8$. These values are used in the $M^{0\nu}$
calculations; see Section \ref{sec4}.

\begin{figure}
\includegraphics[width=2.5in]{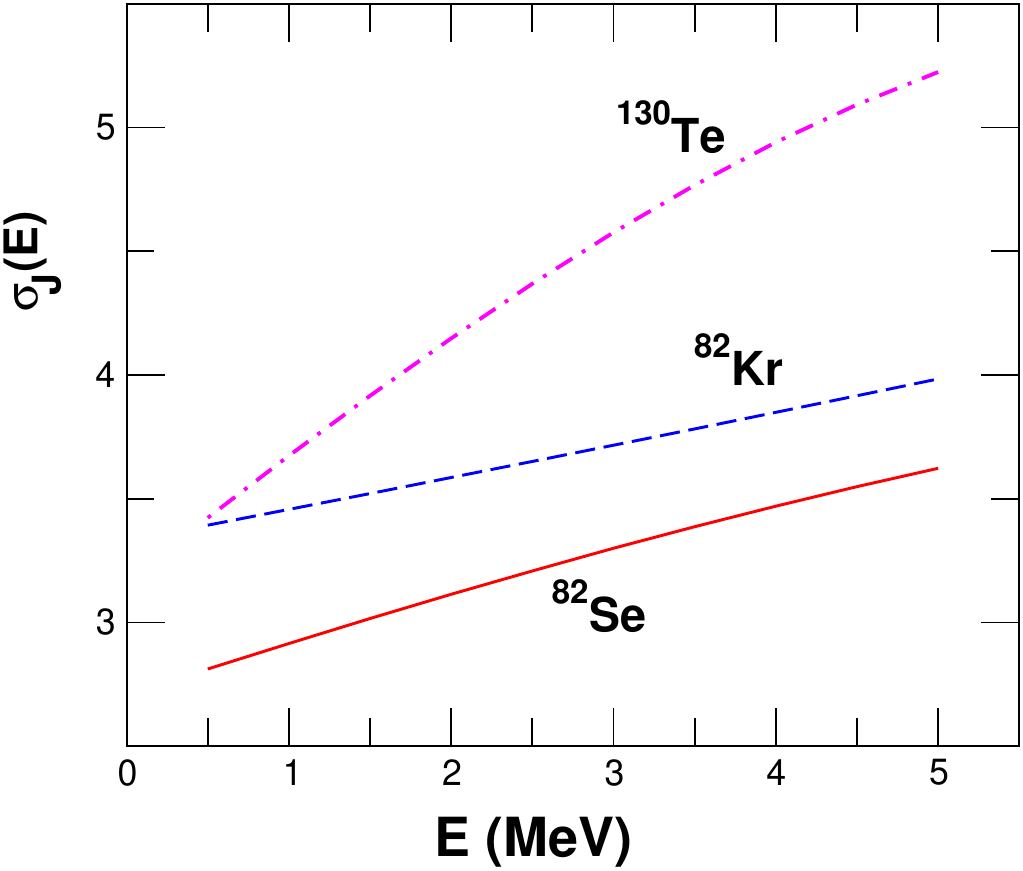}
\caption{SSM results for spin-cutoff factor $\sigma_J(E)$ vs $E$ for $^{82}$Se,
$^{82}$Kr and $^{130}$Te. Details of the sp orbits, spe, and TBME employed in 
the calculations are given in Section \ref{sec4}. Results for $^{130}$Te and 
$^{82}$Se are reported before in \cite{KC-2018}.}
\label{fig1}
\end{figure}

In order to apply Eq. (\ref{eq.ndbd3}), in addition to the marginal centroids, variances and $\zeta$, we also need an expression for $\l|\lan \wm_f \mid \co \mid \wm_i\ran\r|^2$, the configuration mean square matrix element of the transition operator.  Applying the propagation theory described in 
\cite{CFT}, we will have the following simple formula,
\be
\barr{l}
\l|\lan (\tmp, \tmn)_f \mid \co(2: 0\nu) \mid (\tmp, 
\tmn)_i \ran\r|^2 = \l\{ d[(\tmp, \tmn)_f]\r\}^{-1} 
\\ 
\times \;\dis\sum_{\alpha^n , \beta^n , \gamma^p ,\delta^p}\;
\dis\frac{m^i_n(\alpha^n)\, [m^i_n(\beta^n)-\delta_{\alpha^n \beta^n}]\, 
[N_p(\gamma^p) - m^i_p(\gamma^p)]\, [N_p(\delta^p) -
m^i_p(\delta^p) - \delta_{\gamma^p \delta^p}]}{N_n(\alpha^n)\, [N_n(\beta^n) -
\delta_{\alpha^n \beta^n}]\, N_p(\gamma^p)\, [N_p(\delta^p) - \delta_{\gamma^p \delta^p}]}
\\ 
\times \dis\sum_{J_0} \; \l[\co^{J_0}_{j^p_{\gamma^p} j^p_{\delta^p} j^n_{\alpha^n}
j^n_{\beta^n}}(0\nu)\r]^2
(2J_0 +1) \;;\\ 
(\tmp, \tmn)_f = (\tmp, \tmn)_i \times \l(1^+_{\gamma^p} 1^+_{\delta^p}
1_{\alpha^n} 1_{\beta^n} \r) \;.
\earr \label{eq.ndbd4}
\ee
Note that in Eq. (\ref{eq.ndbd4}), the final configuration is defined by removing  one neutron from orbit $\alpha^n$ and another from $\beta^n$ and then adding one proton in orbit $\gamma^n$ and another in orbit $\delta^n$. Also, $N_p(\gamma^p)$ is the degeneracy of the proton orbit $\gamma^p$ and similarly
$N_n(\alpha^n)$ for the neutron orbit $\alpha^n$. Going further,  
for the NTME calculations, we need $J$ projection in $\l|\lan E_f \mid \co \mid E_i\ran\r|^2$ as the quantity of interest is 
$$
\l|\lan E_f J_f^\pi = 0^+\mid \co \mid E_i J_i^\pi = 0^+\ran\r|^2\;.
$$
Here $E_i$ and $E_f$ are the ground state energies of the parent and daughter nuclei respectively and similarly $J_i$ and $J_f$. As we have in our applications only even-even nuclei, the ground states have always $J^\pi=0^+$. 
Using Bethe spin-cutoff factor approximation in SSM \cite{CFT}, we have
\be
\barr{l} 
\l|\lan E_f J_f^\pi = 0^+ \mid \co(2: 0\nu) \mid E_i J_i^\pi 
= 0^+ \ran\r|^2  =
\dis\frac{\l|\lan E_f \mid \co(2:0\nu) \mid E_i \ran\r|^2}{\dis\sqrt{
C_{J_i=0}(E_i) C_{J_f=0}(E_f)}}\; \;; \\ 
\\
C_{J_r}(E_r)=\dis\frac{(2J_r+1)}{\dis\sqrt{8\pi}\; \sigma^3_J(E_r)}  \exp
-[(2J_r+1)^2/8\sigma_J^2(E_r)] \stackrel{J_r=0}{\longrightarrow} \dis\frac{1}
{\dis\sqrt{8\pi} \sigma^3_J(E_r)} 
\earr \label{eq.ndbd5} 
\ee
where $r=i,f$. Note that $\sigma_J^2(E) = \lan J_Z^2\ran^E$ is the energy
dependent spin-cutoff factor. Note that in the last step in
Eq. (\ref{eq.ndbd5}) used is the fact that in general $\sigma_J(E) >> 1$. The energy dependent spin-cutoff factor can be calculated using SSM (without considering the dependence on the proton-neutron configurations) starting with the shell model sp orbits, spe, TBME and $(m_p,m_n)$ \cite{Fr-06}. This is applied to some nuclei of interest and the results are shown in Fig. \ref{fig1}. It is seen that $\sigma_J(E) \sim 3-6$ with $E$ varying up to 5 MeV excitation. In the present work $M^{0\nu}$ is calculated using
Eq. (\ref{eq.ndbd5}) by varying $\sigma_J$ from 3 to 6.  Finally, we need to determine the ground state of the
parent and daughter nuclei as Gaussians extend to $\infty$ in Eq. (\ref{eq.ndbd1}). For determining the ground state with energy $E_{gs}$ we use the so called Ratcliff procedure \cite{Ratcliff,CFT,KH}. This needs a reference energy level (from data) with its excitation energy ($E_R$) and $J^\pi$ value ($J_R^\pi$) and also the total
number of states ($N_R$ - this includes $(2J+1)$ factor) up to and including the reference state. The constraint
here being that the $J^\pi$ values for all the levels below the reference level should be known firmly. Then the ground state is given by inverting (numerically) the equation
\be
N_R-[(2J_R+1)/2] = \dis\sum_{(\tmp,\tmn)}\,d(\tmp,\tmn)\,\dis\int_{-\infty}^{E_{gs}+E_R}\;\rho_{\cg}^{(\tmp, \tmn)}(E)\,dE\;.
\label{eq.ndbd6}
\ee   
Eqs. (\ref{eq.ndbd1})-(\ref{eq.ndbd6}) along with Eqs. (\ref{eq.dbd2})-(\ref{eq.dbd4}) will allow one to obtain NTME values using SSM.

\begin{table}
\caption{Values of various parameters in SSM calculations. See text for the description of the entries in each column in the table. The data values for $(E_R , J_R^\pi , N_R)$ are from \cite{nndc}.}
\label{tab1}
\begin{tabular}{c|c|c|c|c|c}
\hline
nucleus & $(m_p,m_n)$ & $N[(\tmp, \tmn);+]$ & $\overline{\sigma(\tmp, \tmn)}$ 
& $\Delta$ (\%) & $(E_R , J_R^\pi , N_R)$ \\
\hline
$^{76}$Ge & $(4,16)$ & 958 & $4.4$MeV & 6\% & $(2.02 \mbox{MeV}, 4^+, 37)$ \\
$^{76}$Se & $(6,14)$ & 2604 & $5.51$MeV & 4\% & $(1.79 \mbox{MeV},\,
2^+,\,33)$ \\
$^{82}$Se & $(6,20)$ & 316 & $3.34$MeV & 9\% & $(1.735 \mbox{MeV},\, 4^+,\,21)$ \\
$^{82}$Kr & $(8,18)$ & 1354 & $4.7$MeV & 5\% & $(2.172 \mbox{MeV},\, 0^+,\,34)$ \\
$^{100}$Mo & $(14,20)$ & 22922 & $6.85$MeV & 2\% & $(1.464 \mbox{MeV},\, 2^+,\,26)$ \\
& & (679) & & & \\
$^{100}$Ru & $(16,18)$ & 21792 & $6.97$MeV & 2\% & $(2.1 \mbox{MeV},\,2^+,\,62)$\\
& & (726) & & & \\
$^{124}$Sn & $(0,24)$ & 188 & $1.94$MeV & 5\% & $(2.192 \mbox{MeV},\, 0^+,\,21)$ \\
$^{124}$Te & $(2,22)$ & 4079 & $3.35$MeV & 5\% & $(2.225 \mbox{MeV},\,4^+,\,76)$\\
$^{130}$Te & $(2,28)$ & 554 & $2.09$MeV & 9\% & $(1.633 \mbox{MeV},\, 4^+,\,20)$ \\
$^{130}$Xe & $(4,26)$ & 5848 & $3.37$MeV & 6\% & $(1.205 \mbox{MeV},\,4^+,\,20)$ \\
$^{136}$Xe & $(4,32)$ & 42 & $1.05$MeV & 12\% & $(1.892 \mbox{MeV},\, 6^+,\,28)$ \\
$^{136}$Ba & $(6,30)$ & 1394 & $2.61$MeV & 7\% & $(2.141 \mbox{MeV},\,0^+,\,41)$ \\
\hline
\end{tabular}
\end{table}
\begin{figure}
\centerline{\includegraphics[width=250pt]{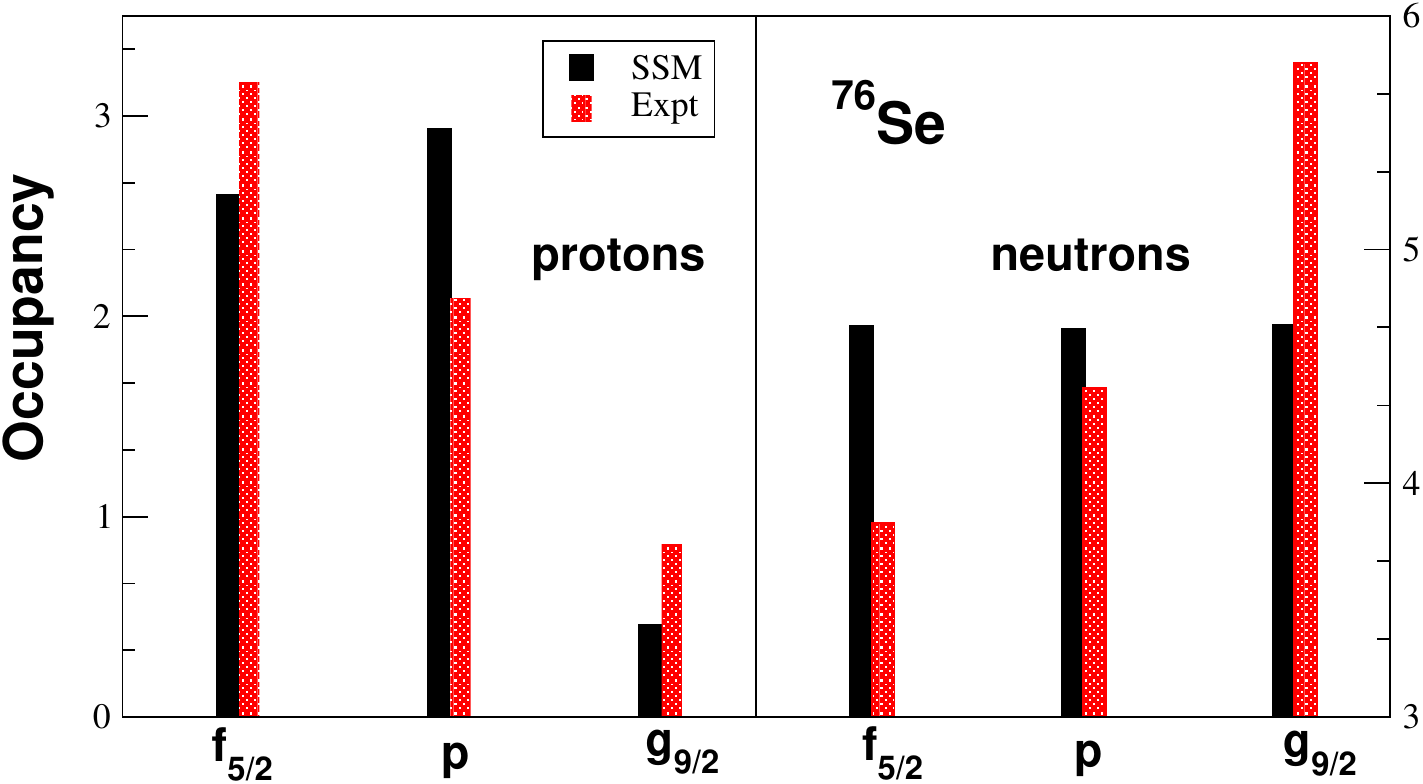}}
\vskip 1cm  
\includegraphics[width=250pt]{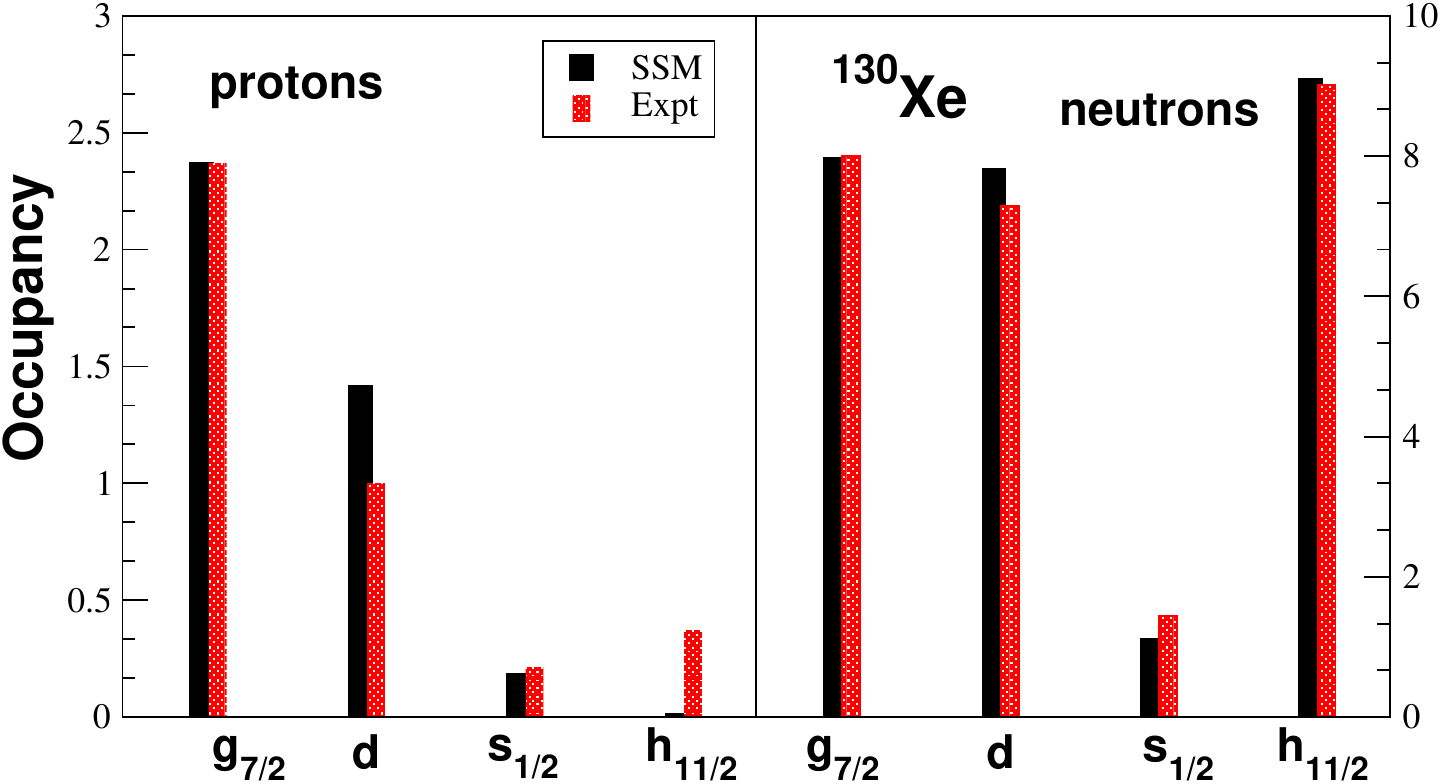}
\caption{SSM results for occupancies for protons and neutrons in $^{76}$Se compared with experimental data \cite{Sch-1,Sch-2}. Note that the occupancies for
$^{1}p_{3/2}$ and $^{1}p_{1/2}$ are summed and shown as the occupancy for the $^1p$ orbit. Similarly, shown also are occupancy results for $^{130}$Xe compared with experimental data \cite{Sch-3,Sch-4}. Here
occupancies for $^{1}d_{5/2}$ and $^{1}d_{3/2}$ are summed and shown as the occupancy for the $^1d$ orbit. Results for $^{130}$Xe were reported earlier in a conference proceedings \cite{Ko-2017}.}
\label{fig2}
\end{figure}
\begin{figure}
\centerline{\includegraphics[width=250pt]{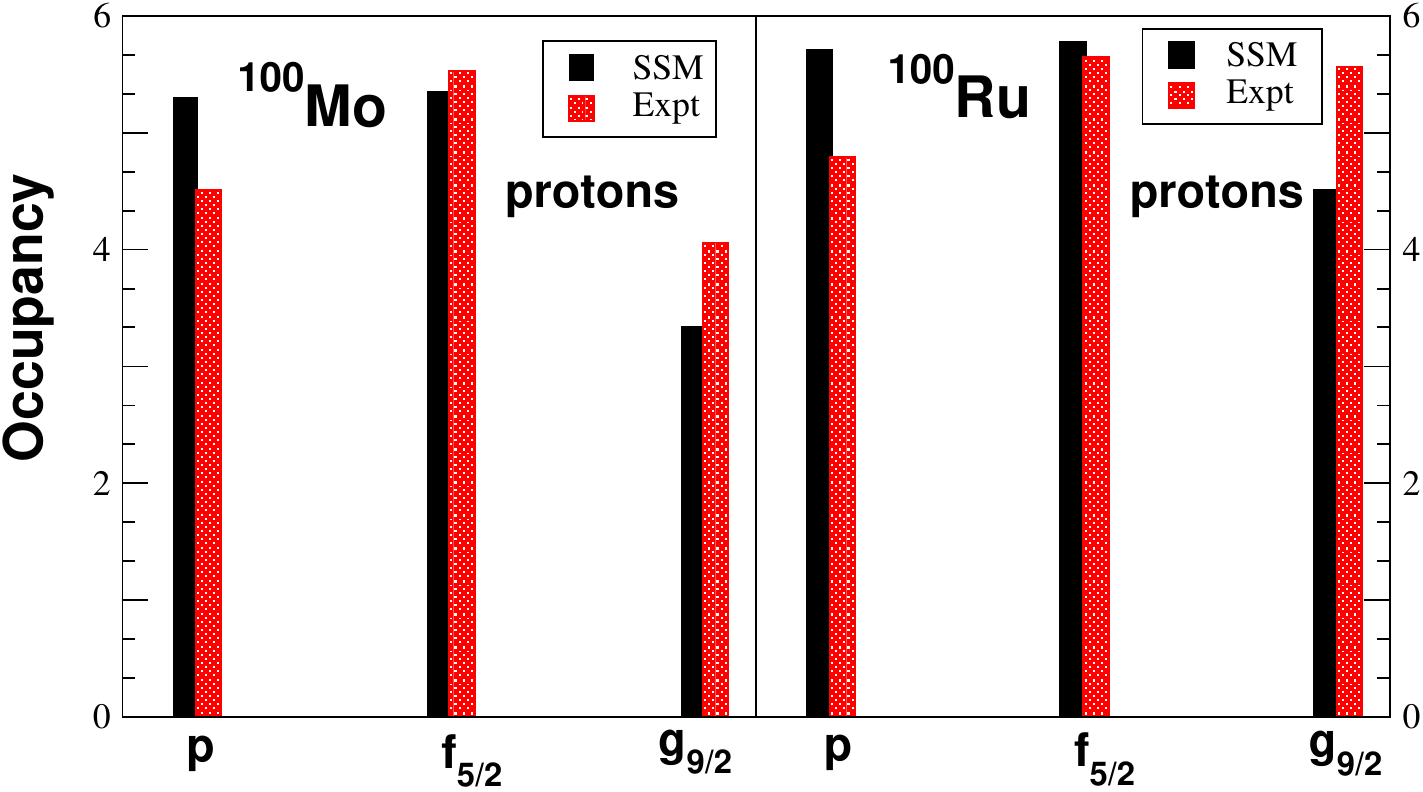}}
\vskip 1cm  
\includegraphics[width=250pt]{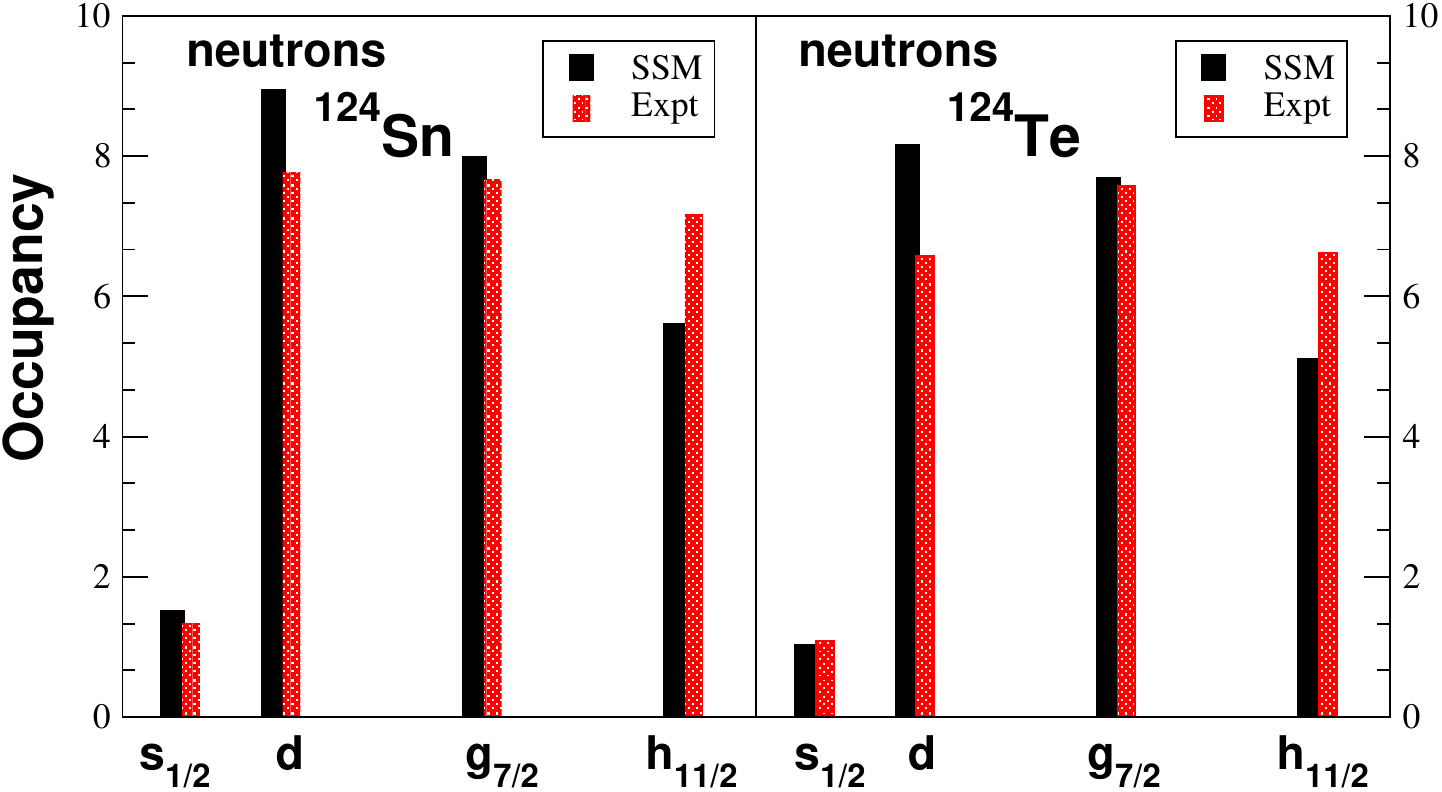}
\caption{Upper panel: SSM results for occupancies for protons in $^{100}$Mo and $^{100}$Ru compared with experimental data \cite{snt-mo100}. Note that the occupancies for
$^{1}p_{3/2}$ and $^{1}p_{1/2}$ are summed and shown as the occupancy for the $^1p$ orbit. Lower panel: SSM results for occupancies for neutrons in $^{124}$Sn ans $^{124}$Te compared with experimental data \cite{snt-sn124}. Here
occupancies for $^{1}d_{5/2}$ and $^{1}d_{3/2}$ are summed and shown as the occupancy for the $^1d$ orbit.}
\label{fig2-new}
\end{figure}

\section{SSM results for $^{76}$Ge, $^{82}$Se, $^{100}$Mo, $^{124}$Sn, $^{130}$Te and $^{136}$Xe NTME} 
\label{sec4}

\subsection{$^{76}$Ge and $^{82}$Se NTME}
\label{sec41}

Neutrinoless double beta decay of $^{76}$Ge to $^{76}$Se is the subject of GERDA \cite{Gerda} and Majorana \cite{Major} experiments while $^{82}$Se to $^{82}$Kr is the subject of CUPID-0 \cite{Cupid0} and NEMO-3 \cite{Nemo3} experiments. In the SSM calculations, $^{56}$Ni is taken as the
core with the valence protons and neutrons  in both $^{76}$Ge and $^{82}$Se (also for the daughter nuclei  $^{76}$Se  and $^{82}$Kr) occupying  the orbits $^1p_{3/2}$, $^0f_{5/2}$, $^1p_{1/2}$ and $^0g_{9/2}$. The effective interaction used is JUN45. The spe (4 in number) and TBME (133 in number)
defining JUN45 are given in \cite{jun45}. Table \ref{tab1} gives 
number ($N[(\tmp, \tmn);+]$) of positive parity proton-neutron  configurations for each nucleus involved. For the partial densities defined over all these
configurations, the exact energy centroids $E_c$ and variances $\sigma^2$ are calculated using trace propagation equations and then the Gaussian partial densities are obtained. In Table \ref{tab1} also shown is the average proton-neutron configuration width 
$\overline{\sigma} = \overline{\sigma(\tmp, \tmn)}$ and also the width ($\Delta$ in \%) of its flucuations over the various configurations. In addition, the parameters $(E_R , J_R^\pi , N_R)$ for determining the ground state are also given in Table \ref{tab1} and this data is obtained from \cite{nndc}. For all above four nuclei, the ground states, obtained using Eq. (\ref{eq.ndbd6}), are found to be $\sim (2.8-3) \overline{\sigma}$  below the lowest configuration  centroid. 

Goodness of the Gaussian partial densities as well as the ground state determination are tested using occupancies of the shell model $j$ orbits. In SSM, occupancy of a $j$ orbit is simply given by \cite{CFT,KP-79},
\be
\lan {\hat{n}}^{\rho}_j\ran^{E_{gs}} = \dis\frac{\dis\sum_{(\tmp,\tmn)}\,m^{\rho}_j(\tmp,\tmn)\;I_{\cg}^{(\tmp, \tmn)}(E_{gs})}{\dis\sum_{(\tmp,\tmn)}\,I_{\cg}^{(\tmp, \tmn)}(E_{gs})}\;; \rho=p\;\;\mbox{or}\;\;n\;.
\label{eq.occu}
\ee
Here, $\hat{n}$ is the number operator. As examples, we show in Fig. \ref{fig2} the occupancy results for $^{76}$Se and $^{130}$Xe for both protons and neutrons. Similarly, we show in Fig. \ref{fig2-new} occupancy results for $^{100}$Mo, $^{100}$Ru, $^{124}$Sn and $^{124}$Te. Details of the shell model space and the effective interaction that are employed
for $^{76}$Se and $^{130}$Xe are given in Sections \ref{sec41} and \ref{sec43} respectively. Similarly, details
of the shell model space and the effective interaction that are employed for the four nuclei in Fig. \ref{fig2-new} are given in Section \ref{sec42}. It is clearly seen from Figs. \ref{fig2} and \ref{fig2-new} that the SSM values for occupancies agree quite well with the experimental data. See \cite{Ko-2017,KC-2018} for comparison between SSM results and experimental data for $^{76}$Ge, $^{130}$Te, $^{136}$Xe and $^{136}$Ba. The data for proton and neutron orbit occupancies for all these nuclei are from \cite{Sch-1,Sch-2,Sch-3,Sch-4,snt-mo100,snt-sn124,Sch-5}.
 
\begin{figure}
\includegraphics[width=4in]{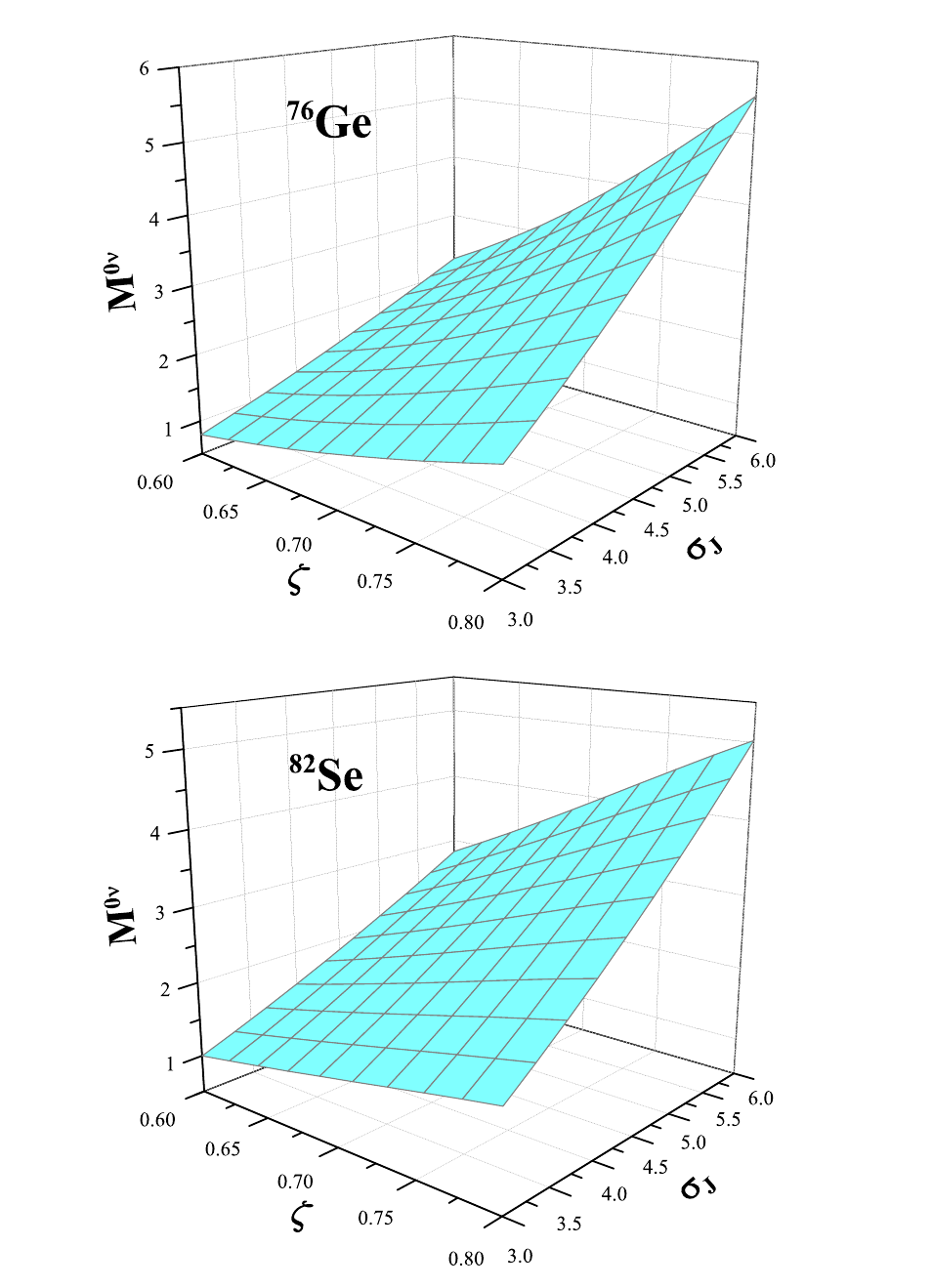}
\caption{SSM results $M^{0\nu}$ vs $(\zeta , \sigma_J)$ for $^{76}$Ge and $^{82}$Se. The figure for $^{76}$Ge is taken 
from \cite{KC-2018}. See text for other details.}
\label{fig3}
\end{figure}

With the ground states determined,
Eqs. (\ref{eq.ndbd1}), (\ref{eq.ndbd3})-(\ref{eq.ndbd5}) and
(\ref{eq.dbd2})-(\ref{eq.dbd4}) are used to calculate 
$M^{0\nu}$. In all the calculations, the bivariate correlation coefficient $\zeta$ is varied between
$0.6$ and $0.8$ and similarly, assuming $\sigma_J(E_i(gs)) =
\sigma_J(E_f(gs))=\sigma_J$ with the $\sigma_J$ value varied between 3 to 6. In Fig. \ref{fig3}, results for $M^{0\nu}$ vs  ($\zeta$, $\sigma_J$) are shown. Following the results
in Fig. \ref{fig1} and using $\sigma_J \sim 3-4$ and similarly, following the results in
\cite{Ko-15} and using $\zeta  \sim 0.7-0.75$, the SSM value for $M^{0\nu} \sim
1.3-2.5$ for $^{76}$Ge as seen from Fig. \ref{fig3}. In comparison,  various shell model calculations (due to groups of Poves, Horoi and Coraggio - see Table I in \cite{Rmp-2023})  give the
value to be in the range $2.66-3.57$. Similarly, for $^{82}$Se, with the same range for $\zeta$ and $\sigma_J$, the SSM
value is $M^{0\nu} \sim 1.4-2.4$ as seen from Fig. \ref{fig3}. In comparison, the shell model values are in the range $2.72-3.39$ \cite{Rmp-2023}.  Let us add that with other nuclear
models such as QRPA, EDF and IBM, the $M^{0\nu} \sim 3-6$ \cite{Rmp-2023}. Thus, it is plausible to conclude that SSM is useful for calculating NTME for $0\nu \beta\beta$. 

\subsection{$^{100}$Mo and $^{124}$Sn NTME}
\label{sec42}

\subsubsection{$^{100}$Mo}

The $^{100}$Mo to $^{100}$Ru NDBD search is the subject of CUPID-Mo experiment \cite{CUPID-Mo} giving a lower bound of $1.8 \times 10^{24}$ yr. Another experiment is AMoRE \cite{Amore}. In the coming years the CUPID experiment is expected to increase this to
$(1-9) \times 10^{27}$yr \cite{CUP-Mo,Barb-23} and similarly, the AMoRE experiment  in future is expected to increase
this limit to $\sim (5-8) \times 10^{26}$ yr \cite{Amore,Barb-23}. Following these, we have carried out SSM calculations for NDBD NTME for $^{100}$Mo.
Recently cross sections of neutral-current neutrino scattering on $^{100}$Mo was studied using DSM \cite{nu-nuc}. In this study employed successfully is the  effective interaction GWBXG with $^{66}$Ni as the closed core. The details of this effective interaction have been 
discussed in \cite{Dey}. Here, the active proton orbits are $^0f_{5/2}$, $^1p_{3/2}$, $^1p_{1/2}$ and $^0g_{9/2}$ with spe
$-5.322$, $-6.144$, $-3.941$ and $-1.250$ MeV respectively. For neutrons, the active orbits are $^1p_{1/2}$, $0g_{9/2}$, $^0g_{7/2}$, $^1d_{5/2}$ and $^1d_{3/2}$ with spe $-0.696$, $-2.597$, $5.159$, $1.830$ and $4.261$ MeV respectively. In our SSM calculations for $^{100}$Mo
NDBD we have employed the same shell model space and the effective interaction; the $^2s_{1/2}$ orbit is not considered just as in \cite{nu-nuc}. Some of the relevant parameters for $^{100}$Mo and $^{100}$Ru are listed in Table \ref{tab1}. Note that the total number of proton-neutron configurations are very large ($\sim 22000$) for these nuclei. However, as shown in the brackets in Table \ref{tab1}, the number of configurations that contribute to the ground states of these nuclei is much smaller ($\sim 700$). For $^{100}$Mo the gs is $3.13\;\overline{\sigma}$  below the lowest configuration centroid and similarly, for $^{100}$Ru the gs is 
$3\;\overline{\sigma}$  below the lowest configuration
centroid. The NTME for $^{100}$Mo to $^{100}$Ru decay are presented in Fig. \ref{fig4} as a functions of ($\zeta$, $\sigma_J$). It is seen from Fig. \ref{fig4} that $M^{0\nu}$ varies from $0.63 - 1.1$ for the choice $\zeta=0.7-0.75$ and $\sigma_J=4-5$. These values are smaller than the shell model value 
of 2.4 from Coraggio et al. \cite{th-sm2}. Other models give the value in the range of 3.9 to 6.59 \cite{Rmp-2023}.

\subsubsection{$^{124}$Sn}

\begin{figure}[ht]

\centerline{\includegraphics[width=250pt]{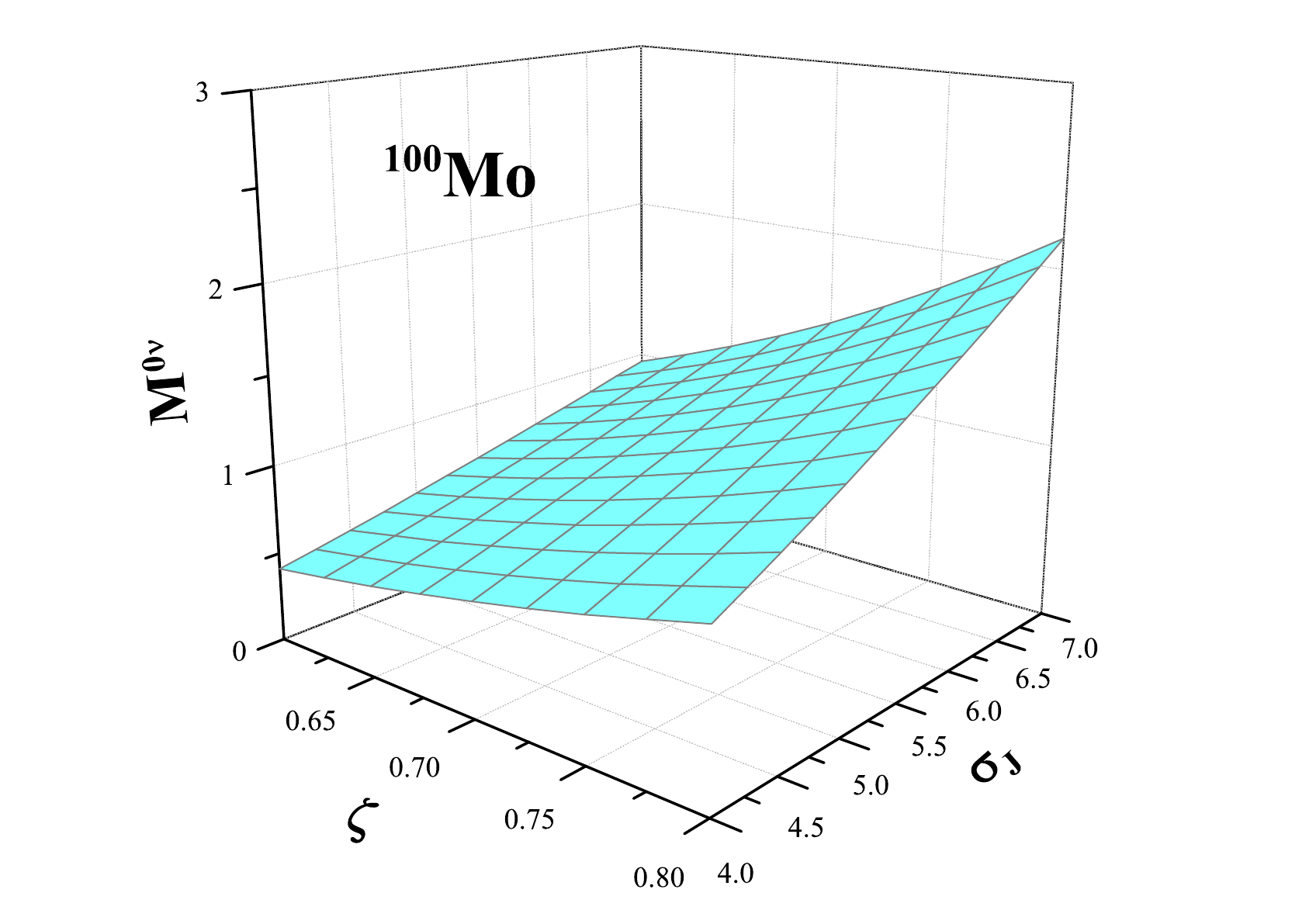}}
\vskip 1cm  
\includegraphics[width=250pt]{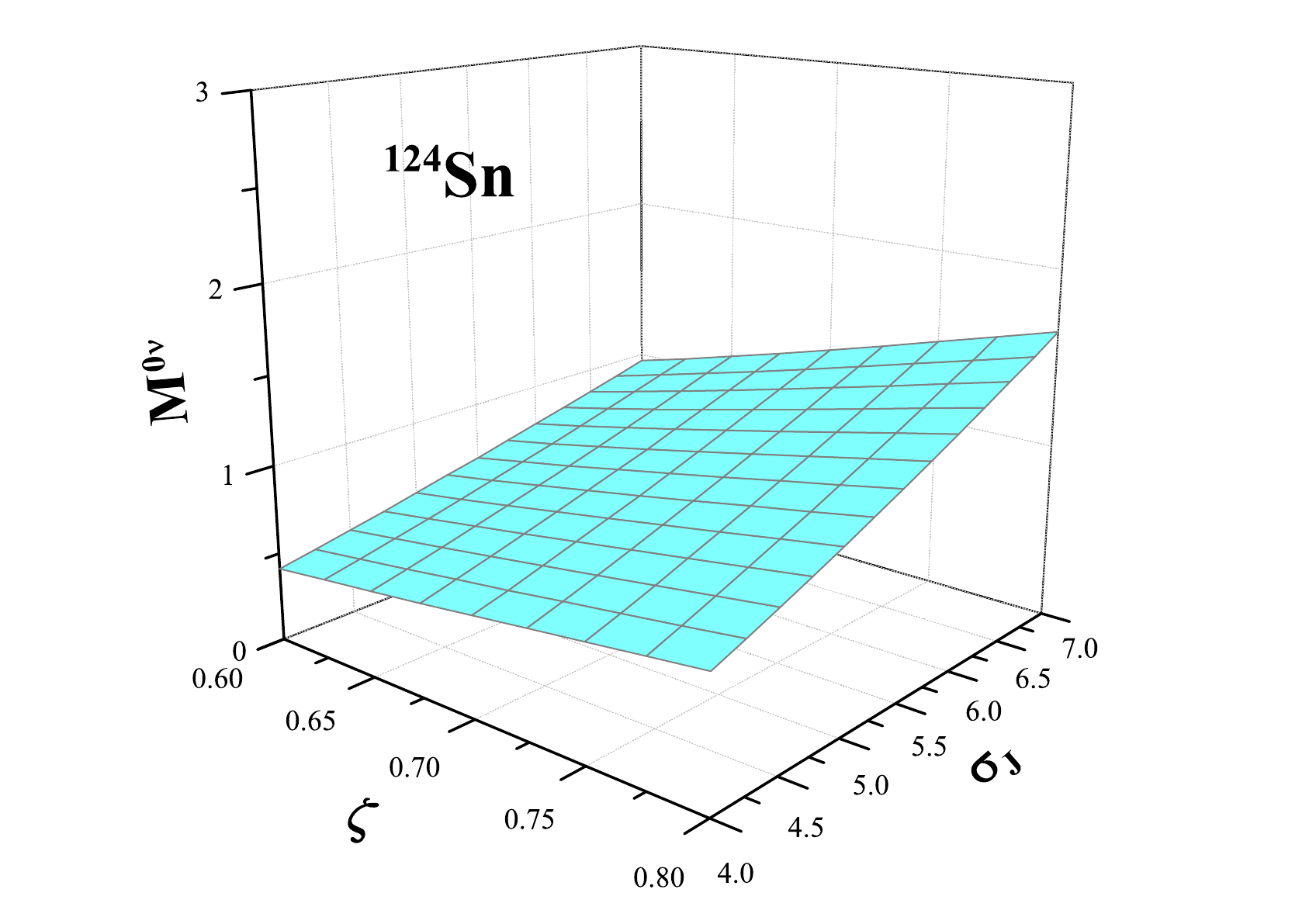}

\caption{SSM results for $M^{0\nu}$ vs $(\zeta , \sigma_J)$
for $^{100}$Mo and $^{124}$Sn. See text for other details.}

\label{fig4}
\end{figure}

The $^{124}$Sn to $^{124}$Te NDBD is of interest in particular in the context of  India based Neutrino Observatory (INO) \cite{Sn124-exp} and therefore we have carried out SSM calculations for the NDBD NTME for
$^{124}$Sn. In the calculations, $^{100}$Sn is taken as the
core with the valence neutrons in $^{124}$Sn and both valence protons and neutrons $^{124}$Te occupying  the orbits $^0g_{7/2}$,
$^{1}d_{5/2}$, $^{1}d_{3/2}$, $^{2}s_{1/2}$ and $^{0}h_{11/2}$. The effective interaction used is jj55. The spe (5 in number) and TBME (327 in number) defining jj55 are given in \cite{jj55-int}.
Some of the relevant parameters for $^{124}$Sn and $^{124}$Te are listed in Table \ref{tab1}. For $^{124}$Sn the gs is 3.3 $\overline{\sigma}$  below the lowest configuration
centroid and similarly, for $^{124}$Te the gs is 2.9 $\overline{\sigma}$  below the lowest configuration
centroid. The NTME for $^{124}$Sn decay to $^{124}$Te decay are presented in Fig. \ref{fig4} as a functions of ($\zeta$, $\sigma_J$). It is seen from Fig. \ref{fig4} that $M^{0\nu}$
varies from $0.6-0.9$ for the choice $\zeta=0.7-0.75$ and $\sigma_J=4-5$. These values are a factor of 2 smaller than the shell model values from Horoi {\it et al.} \cite{Sn124} and Poves {\it et al.} \cite{Poves}.

\subsection{$^{130}$Te and $^{136}$Xe NTME}
\label{sec43}

\begin{figure}[ht]

\centerline{\includegraphics[width=250pt]{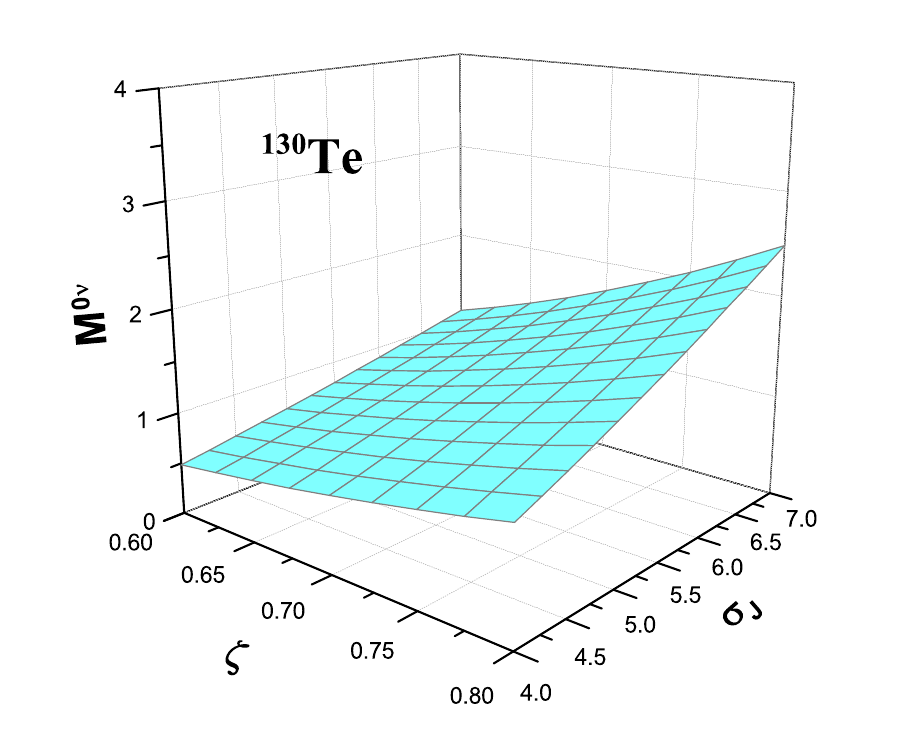}}
\includegraphics[width=250pt]{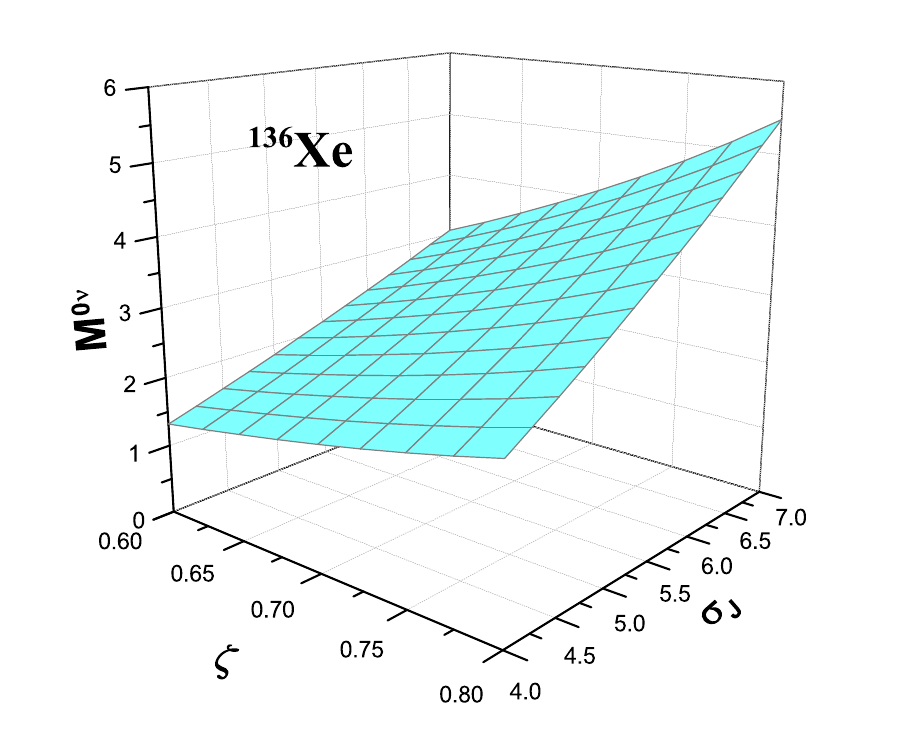}

\caption{SSM results for $M^{0\nu}$ vs $(\zeta , \sigma_J)$ for $^{130}$Te and 
$^{136}$Xe.The figures for both $^{130}$Te and 
$^{136}$Xe appeared earlier in a conference proceedings \cite{Ko-2017}; 
the figure for $^{136}$Xe is also given in \cite{KC-2018}. 
See text for other details.}

\label{fig5}
\end{figure}

Neutrinoless double beta decay of $^{130}$Te to $^{130}$Xe is the subject of CUORE \cite{Curoe} experiment while $^{136}$Xe to $^{136}$Ba is the subject of KamLAND-Zen \cite{Kamland} and EXO-200 \cite{Exo-200} experiments. In the SSM calculations, the model space and the effective interaction used for $^{130}$Te and $^{136}$Xe (also for the daughter nuclei $^{130}$Xe and $^{136}$Ba) are same as those described for $^{124}$Sn. Table \ref{tab1} gives number $N[(\tmp, \tmn);+]$ and some other parameters for the four nuclei $^{130}$Te, $^{130}$Xe, $^{136}$Xe and $^{136}$Ba. For $^{130}$Te the gs is 2.9$\overline{\sigma}$ below the lowest configuration centroid. Similarly, for $^{130}$Xe the gs is
3.2$\overline{\sigma}$  below the lowest configuration centroid. 
With this, NTME are calculated for $^{130}$Te decay to $^{130}$Xe and the
results are presented in Fig. \ref{fig5} as a function of ($\zeta$, $\sigma_J$) with $\zeta$ varying from 0.6 to 0.8 and $\sigma_J$ from 4 to 7. Following the results in Fig. \ref{fig1} and using $\sigma_J \sim 4-5$ and similarly, following the results in \cite{Ko-15} and using $\zeta  \sim 0.7-0.75$, the SSM value for $M^{0\nu} \sim 0.8-1.3$ for $^{130}$Te as seen from Fig. \ref{fig5}. In comparison, the shell model values vary from 1.79 to 3.16 and the QRPA, EDF and IBM values vary in the range 1.4 to  6 \cite{Rmp-2023}.  
Turning to $^{136}$Xe NDBD,  It is seen from Fig. \ref{fig5} that $M^{0\nu}$ varies from $1.8-2.5$ for the choice $\zeta=0.7-0.75$ and $\sigma_J=4-5$. Shell model values for $^{136}$Xe vary from 1.63 to 2.45 while the QRPA, EDF and IBM values vary from 1.11 to 4.7 \cite{Rmp-2023}. 

\subsection{Summary}
\label{sec44}

\begin{figure}
\includegraphics[width=4in,angle=0]{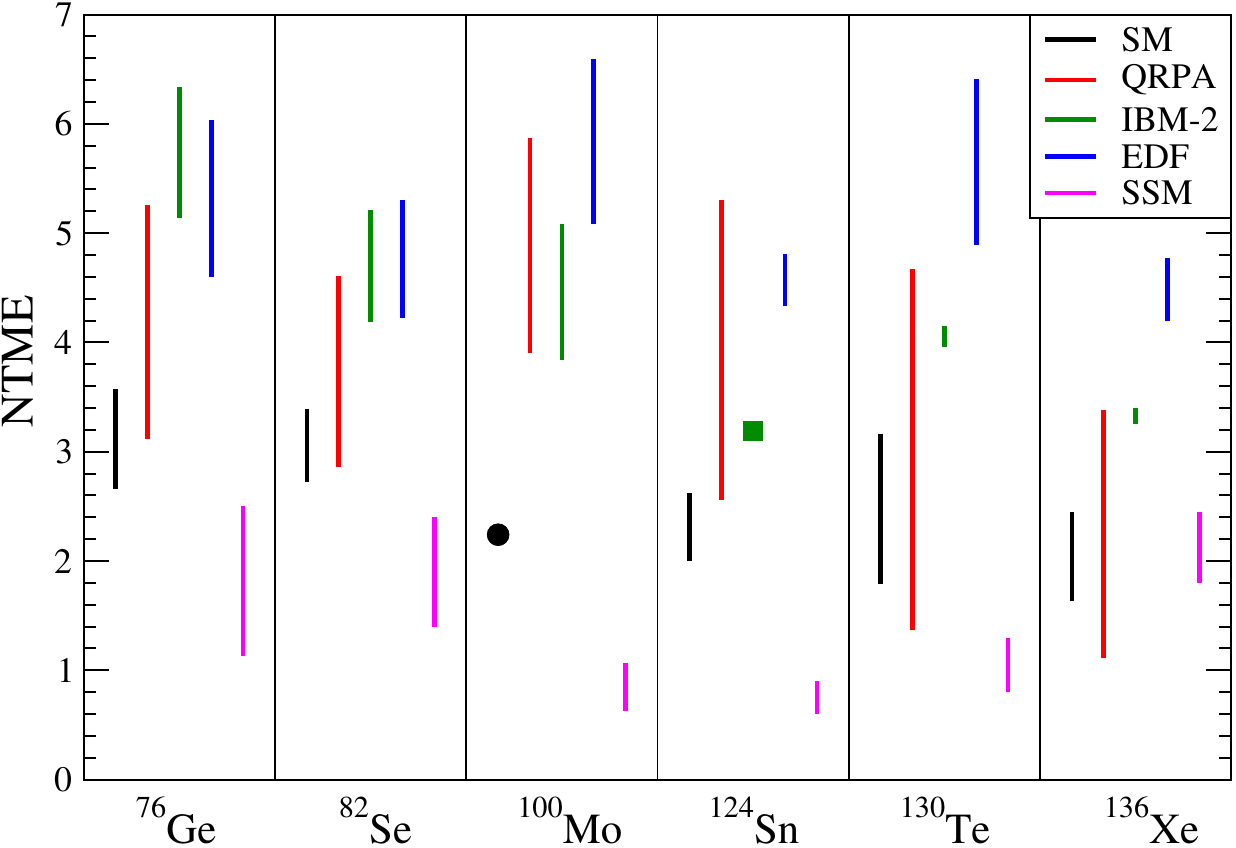}

\caption{Results for NTME (i.e. $M^{0\nu}$) from shell model (SM), QRPA, IBM-2 and EDF theory compared with the SSM results described in Section \ref{sec4}. There is  variation in SSM results due to a range of $\zeta$ and $\sigma_J$ values used in the calculations. Similarly, the variation in other models is due to the fact that there are several calculations using the same model. Results for SM are taken from \cite{th-sm1,th-sm2,th-sm3,th-sm4,Sn124,Rmp-2023}, for QRPA from \cite{th-qrpa1,th-qrpa2,th-qrpa3,Sn124,Rmp-2023}, for IBM-2 from \cite{th-iac1,th-iac2,Sn124,Rmp-2023} and for EDF from \cite{th-edf1,th-edf2,Sn124,Rmp-2023}. For each nucleus, results are shown for SM, QRPA, IBM-2, EDF and SSM in that order. Note that for $^{100}$Mo there is only one SM calculation and similarly for $^{124}$Sn there is only one IBM-2 calculation.}

\label{fig6}
\end{figure}

We have presented SSM results for NTME in Figs. \ref{fig3}, \ref{fig4} and \ref{fig5}. These results are compared with the results from other nuclear structure models in Fig. \ref{fig6}. As it is well known,  results from QRPA, EDF and IBM-2 are in general larger than SM results. As seen from the figure, results from the present statistical method (SSM) are within a factor of 2 compared to the shell model results. 
This shows that the results from SSM might improve further if one is able to relax some of the approximations used in Section III.

\section{Conclusions and future outlook}
\label{sec5}

Statistical shell model method for NDBD NTME $M^{0\nu}$ calculations is applied to  $^{76}$Ge, $^{82}$Se, $^{100}$Mo, $^{124}$Sn, $^{130}$Te and $^{136}$Xe nuclei that are of current experimental interest. Structure of NDBD operator that gives $M^{0\nu}$ is described in Section \ref{sec2}. Similarly, various SSM formulas for calculating NTME are given in Section \ref{sec3}. The SSM results for NTME are presented in Section \ref{sec4}. These results are compared with the results from other nuclear structure models in Fig. \ref{fig6}. It is important to emphasize that SSM is a statistical approach and  recently, Horoi et al. developed a different type of statistical shell model approach for NTME \cite{Horoi-1,Horoi-2}.

In future, as the present SSM results are a factor 2 smaller than shell model results (see Fig. \ref{fig6}), it is important to consider the following and improve the SSM formulation given in Section \ref{sec3}. (i) Evaluate configuration centroids and  variances over fixed-$J$ spaces and with this one need not use spin-cutoff factors. This is indeed possible using for example the large scale computer  codes developed recently by R.A. Sen'kov et al \cite{Zel-1,Zel-3} and extending them to larger spaces considered for $^{100}$Mo, $^{124}$Sn, $^{130}$Te and $^{136}$Xe. In the fixed-$J$ formalism, we also need the formula for
$$
\l|\lan (\tmp, \tmn)_f , J_f=0 \mid \co(2: 0\nu) \mid (\tmp, 
\tmn)_i , J_i=0 \ran\r|^2 \;.
$$
As $\co(2: 0\nu)$ is two-body in nature, in principle it is possible to derive the formula following the methods described in \cite{Wong}. This will be addressed in  future. In addition, treatment of the bivariate correlation coefficient as a function of the initial and final configurations (and also including $J_i$ and $J_f$) need to be developed and this requires further studies using embedded random matrix ensembles. (ii) Another gap in SSM is that the Gaussian forms employed in the theory do not take into account pairing effects adequately. This follows from the fact that pairing generates skewed (Gaussians are symmetric) partial densities \cite{sdm-p1,sdm-p2,sdm-p3}. Also, it is well known that pairing enhances NDBD-NTME values \cite{p-pair,p-deform}. Thus, inclusion of pairing effects in the Gaussian distributions that appear in SSM are expected to enhance the NTME values shown in Fig. \ref{fig6}. (iii) In addition to considering fixed-J averages and including properly pairing effects, it is important to perform future SSM calculations using $q$-normal and bivariate $q$-normal forms following some recent developments described in \cite{KM-qnorm}.  

It is useful to add that Eq. (\ref{eq.ndbd4}) easily gives the total transition  strength sum, the sum of the strengths from all states of the parent nucleus to all the states of the daughter nucleus generated by $\co(2: 0\nu)$ and this depends only on the sp space considered. For example, the $^{82}$Se the total strength sum is 31239 and for $^{76}$Ge it is 54178. Thus, $M^{0\nu}(0^+)$ is a very small fraction of the
total strength generated by the NDBD transition operator. Starting from Eq. (\ref{eq.ndbd3}), it is also possible to obtain the linear and quadratic energy weighted strength sums. These may prove to be useful in putting constraints on the nuclear models that are being used for NDBD studies.

In conclusion, we have presented a first comprehensive set of SSM results for NDBD-NTME and it is our hope that this will generate new interest in developing a more refined SSM approach relaxing various approximations used in the present paper [i.e. by considering (i), (ii) and (iii) above].

\section*{Acknowledgments} VKBK is thankful to R. Haq for initial collaboration and for many useful discussions. Thanks are due to N.D. Chavda for preparing some of the figures.


\ed